\documentclass[aps,prx,twocolumn,superscriptaddress]{revtex4-2}

\usepackage{times}
\usepackage{braket}
\usepackage{xcolor}
\usepackage{graphicx}
\usepackage{multirow}
\usepackage{amssymb}
\usepackage{amsmath}
\usepackage{comment}
\usepackage{csquotes}
\usepackage[normalem]{ulem}

\usepackage{hyperref}
\hypersetup{
    colorlinks=true,
    citecolor=blue,
    urlcolor=blue
}

\def\bibcommenthead{}

\begin{document}

\title{Statistical Localization in a Rydberg Simulator of $U(1)$ Lattice Gauge Theory}

\author{Prithvi Raj Datla}
\affiliation{Centre for Quantum Technologies and Department of Physics, National University of Singapore, 117542 Singapore, Singapore}
\affiliation{Department of Physics, Stanford University, Stanford, CA, USA}
\author{Luheng Zhao}
\affiliation{Centre for Quantum Technologies and Department of Physics, National University of Singapore, 117542 Singapore, Singapore}
\affiliation{Duke Quantum Center, Duke University, Durham, NC 27701, USA}
\affiliation{Department of Electrical and Computer Engineering, Duke University, Durham, NC 27708, USA}
\author{Wen Wei Ho}
\affiliation{Centre for Quantum Technologies and Department of Physics, National University of Singapore, 117542 Singapore, Singapore}
\author{Natalie Klco}
\affiliation{Duke Quantum Center, Duke University, Durham, NC 27701, USA}
\affiliation{Department of Physics, Duke University, Durham, NC 27708, USA}
\author{Huanqian Loh}
\email[]{huanqian.loh@duke.edu}
\affiliation{Centre for Quantum Technologies and Department of Physics, National University of Singapore, 117542 Singapore, Singapore}
\affiliation{Duke Quantum Center, Duke University, Durham, NC 27701, USA}
\affiliation{Department of Electrical and Computer Engineering, Duke University, Durham, NC 27708, USA}
\affiliation{Department of Physics, Duke University, Durham, NC 27708, USA}

\date{\today}
\begin{abstract}
Lattice gauge theories (LGTs) provide a framework for describing dynamical systems ranging from nuclei to materials. LGTs that host concatenated conservation laws can exhibit Hilbert space fragmentation, where each subspace may be labeled by a conserved quantity with nonlocal operator support. It is expected that nonlocal conservation laws will not impede thermalization locally. However, this expectation has recently been challenged by the notion of statistical localization, wherein particular motifs of microscopic configurations may remain frozen in time due to strong Hilbert space fragmentation. Here, we report the first experimental signatures of statistically-localized behavior. We realize a novel constrained LGT model using a facilitated Rydberg atom array, where atoms mediate the dynamics of electric charge clusters whose nonlocal pattern of net charges remains invariant. By experimentally reconstructing observables sampled from a temporal ensemble, we probe the spatial distribution of each conserved quantity. We find that as a result of strong Hilbert space fragmentation, the expectation values of all conserved quantities remain locally distributed in typical quantum states, even though they are described by nonlocal string-like operators. Our work opens the door to high-energy explorations of cluster dynamics and low-energy studies of strong zero modes that persist in infinite-temperature topological systems.

\end{abstract}
\maketitle

Lattice gauge theories (LGTs) form a basic framework for understanding a wide variety of physical systems ranging from the Standard Model of particle physics to condensed matter systems \cite{kogut1983lattice,greensite2003confinement,sachdev2016emergent}. Quantum simulators that realize kinetic constraints \cite{kohlert2023exploring,adler2024observation,kim2023realization,zhao2025observation,honda2025observation,gonzalez2024observation}, like those of strongly-interacting Rydberg atoms, have allowed for a natural way to realize and study a myriad of LGTs experimentally. Often, their quantum dynamics can be naturally restricted to a desired subspace, such as the gauge-invariant subspace obeying Gauss' law \cite{Tagliacozzo:2012df,Banerjee:2012pg,Banuls:2019bmf,Zohar:2012xf,Banuls:2019bmf,schweizer2019floquet,zhou2022thermalization,halimeh2025cold,surace2020lattice,gonzalez2024observation,Bauer:2023qgm}. At the same time, kinetic constraints can lead to slow dynamics, leading to a connection between LGTs and nonergodic quantum phenomena like many-body localization (MBL) or scarring \cite{nandkishore2015many,schreiber2015observation,bernien2017probing,turner2018weak,smith2018dynamical,karpov2021disorder,banerjee2021quantum,aramthottil2022scar,halimeh2023robust}.

An LGT quantum simulator with local kinetic constraints may undergo Hilbert space fragmentation \cite{sala2020ergodicity,khemani2020localization,yang2020hilbert}. For instance, the gauge-invariant subspace may be subdivided into several symmetry sectors corresponding to global conserved charges, following which short-range kinetically-constrained interactions further break a given symmetry sector into Krylov subspaces. Each Krylov subspace may in turn be labeled by conserved operators with nonlocal support. In the limit of strong fragmentation, the largest Krylov subspace occupies a negligible fraction of the entire Hilbert space \cite{sala2020ergodicity,moudgalya2022quantum,ciavarella2025generic}.

In general, the presence of a set of nonlocal conservation laws, even if extensive, is not expected to impede thermalization in a many-body system \cite{deutsch1991quantum,rigol2008thermalization}. However, it has been recently predicted that in the limit of strong fragmentation, almost all states remain either partially or completely localized despite the nonlocal operators characterizing each subspace --- a phenomenon known as statistical localization \cite{rakovszky2020statistical}. This is in contrast to MBL, which has thus far been construed to be described by quasilocal conservation laws \cite{chandran2015constructing,imbrie2017local,rademaker2017many,singh2021local,ros2015integrals,bertoni2024local,wahl2020local}. In statistical localization, \textit{the expectation values} of these conserved operators in \textit{typical} quantum states pick up contributions from a vanishingly small fraction of the system and appear local. Direct experimental signatures of statistical locality have so far remained elusive due to challenges in probing typical quantum states and in resolving the complex nonlocal conserved quantities of these fragmented systems.

\begin{figure*}[!htbp]
    \centering
    \includegraphics[keepaspectratio,width=18cm]{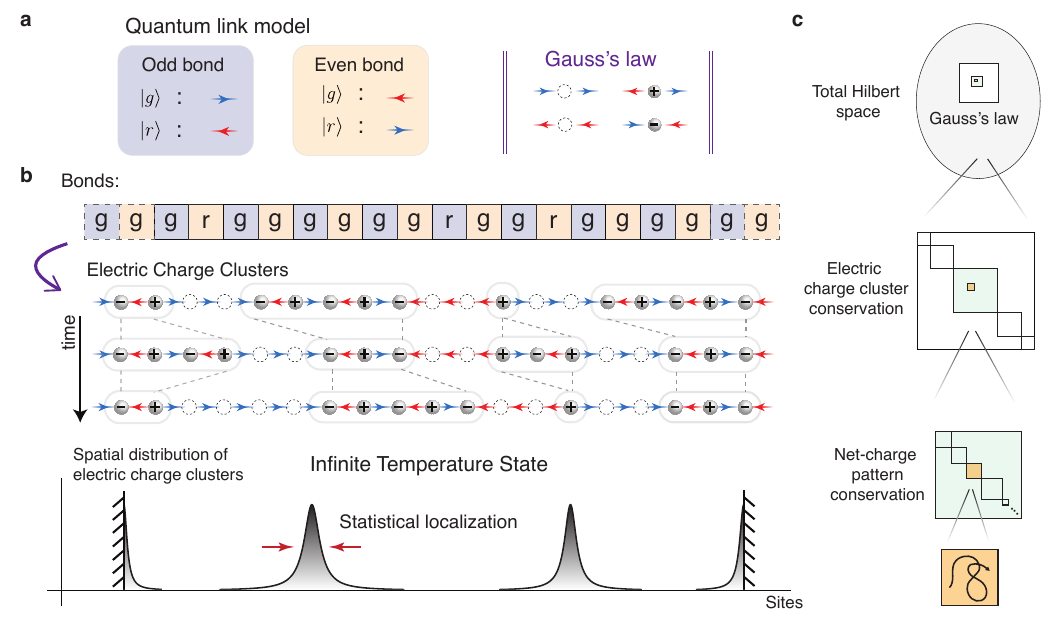}
        \caption{\textbf{Emergence of statistical localization from charge cluster conservation in a constrained $U(1)$ LGT.} \textbf{a}, Representations in the quantum link model. Electric strings on the bonds are described by $\ket{g},\ket{r}$ atoms. Electric charges are placed between the electric strings in a configuration that is consistent with Gauss's law. \textbf{b}, A Rydberg-atom chain, mapped onto electric charge clusters spaced by vacua. The constrained LGT dynamics allow the charge clusters to grow and shrink as long as the net-charge pattern of all the clusters is preserved. Notably, the statistically averaged distribution remains spatially localized in an infinite-temperature state. \textbf{c}, Hilbert space structure of the constrained LGT. The gauge invariance arising from Gauss's law reduces the dimension of the Hilbert space, within which a global conservation of the number of charge clusters $N_{c}$ results in a block-diagonal structure of symmetry sectors, $\mathcal{H}_{N_c}$. Within each symmetry sector, the dynamics are constrained to a Krylov fragment characterized by the pattern of net charges, $\{q_k\}$.} 
 \label{fig:1}
\end{figure*}

In this work, we report the first observation of statistical localization emerging from nonlocal conserved quantities in a $U(1)$ LGT Rydberg simulator, where the Rydberg and ground-state atoms on the bonds mediate the dynamics of electric charge clusters on the sites. Our results are enabled by advances on three fronts. First, we identify a constrained LGT model that yields a microscopic description of statistical localization. We find a natural interpretation for Hilbert space fragmentation in terms of conserving the net-charge pattern of electric charge clusters. Second, we propose a novel effective Hamiltonian with facilitated Rydberg atom arrays for studying the constrained LGT dynamics, which we then experimentally realize. Finally, we demonstrate an efficient protocol for experimentally reconstructing observables for an infinite-temperature state based on sampling from different Krylov subspaces. With these advances, we find signatures of localization that persist even at infinite temperature, despite the conserved quantities having nonlocal operator support.

\section*{Realization of a constrained LGT model}
\textit{Effective models with Rydberg atom arrays.}
Our experiments are performed on one-dimensional Rydberg atom arrays, where regularly spaced $^{87}\text{Rb}$ atoms encode two-level quantum systems in their ground state $\ket{g}$ and Rydberg state $\ket{r}$. The Hamiltonian is given by ($\hbar = 1$):
\begin{equation} \label{eqn:RydHamiltonian}
H_{\text{Ryd}} = -\Delta \sum_{i=1}^{N_a} Q_{i} + \frac{\Omega}{2}\sum_{i=1}^{N_a} X_{i} + \sum_{j = 1}^{2} \sum_{i = 1}^{N_a-j} V_{j-1} Q_{i} Q_{i+j} \, ,
\end{equation}
where $i$ indexes the atom, $\Delta$ is the global detuning, $\Omega$ is the Rabi frequency, $X_{i} = \ket{g_{i}}\bra{r_{i}} + \ket{r_{i}}\bra{g_{i}}$, and $Q_{i} = \ket{r_{i}}\bra{r_{i}}$. The van der Waals interaction between Rydberg atoms, $V_{j-1} = C_6/(ja)^{6}$ for interatomic spacing $a$, is truncated beyond the first two orders (nearest-neighbor $V_{0}$ and next-nearest-neighbor $V_{1}$) due to the approximation $V_{0}, V_{1} \gg \Omega \gg V_{2}$. 

Upon setting $\Delta = V_1$, we realize the ``PPXPQ + QPXPP'' effective Hamiltonian to leading order, where $P = \ket{g}\bra{g}$ is the projector onto the ground state. In other words, an atom is constrained to flip its spin only when both its nearest neighbors are in the ground state and one (but not both) of its next-nearest neighbors is in the Rydberg state. To ensure that such spin-flip behavior is well-defined even at the boundaries of the $N_{a}$ physical array, we introduce ``\emph{$g$-padding}", where two fictitious ground-state atoms are added on each end of the array. We henceforth work with the effective chain of length $N = N_a + 4$.

\textit{Mapping onto a constrained LGT model.}
The effective PPXPQ + QPXPP model exhibits strong Hilbert space fragmentation within the blockaded subspace corresponding to no nearest-neighbor Rydberg excitations. To provide a physical picture for the fragmentation, we map the effective model onto a staggered $U(1)$ lattice gauge theory. Each configuration of atoms maps to a configuration in the quantum link model\cite{surace2020lattice}, where atoms define gauge fields residing on the bonds of the latter chain. Precisely, for odd bonds, an electric field string pointing to the right (left) is mapped from an atom in the ground (Rydberg) state and vice versa for even bonds (Fig.~\ref{fig:1}a). Depending on the relative orientation of two neighboring electric strings, we then allow the sites between the bonds to be occupied by positive charges, negative charges or vacuum, in a way that is consistent with Gauss's law (Fig.~\ref{fig:1}b). Finally, we express the charges and vacua as spin-1/2 degrees of freedom \{$\ket{\circ} ,\ket{\bullet}$\} on matter sites with a compensated staggered mapping (Extended Data Fig.~\ref{fig:ext_fig_mapping}) to allow a site-wise translationally invariant Hamiltonian:
\begin{equation}\label{eqn:LGTHamiltonian}
    H_{\text{LGT}} = -w \sum_{j=2}^{N-3} \Pi_{j-1,j+2} \,\sigma^{+}_j S^{-}_{(j,j+1)} \sigma^{-}_{j+1} + \text{H.c.} \, ,
\end{equation}
where $j$ indexes the matter sites, $\Pi_{j-1,j+2} = \ket{\circ_{j-1}\circ_{j+2}}\bra{\circ_{j-1}\circ_{j+2}} + \ket{\bullet_{j-1}\bullet_{j+2}}\bra{\bullet_{j-1}\bullet_{j+2}}$ projects onto the same-site configuration at $j-1$ and $j+2$, and $\sigma^{+}_{j}$ ($\sigma^{-}_{j}$) is the raising (lowering) operator for a spin on the matter site. $S^{-}_{(j, j+1)}$ acts on the bond between sites $j$ and $j+1$. Explicitly, in terms of the Rydberg degrees of freedom,
$S^{-} = \ket{\rightarrow}\bra{\leftarrow}$. 

In this $U(1)$ quantum link model, blocks of consecutive $\ket{g}$ atoms correspond to electric charge clusters, whereas $\ket{grgrgr\ldots}$ blocks correspond to vacuum clusters separating the charge clusters. The charge clusters may either carry a net charge of $\pm 1$ or be charge neutral. Their dynamics under $H_\text{LGT}$ are such that the complete pattern of net charges is conserved (see Methods), where the net-charge pattern corresponds to a particular Krylov subspace. In other words, the Hilbert space fragmentation takes on a physical meaning in the $U(1)$ LGT: a state initialized with a given net-charge pattern cannot evolve into a state with a different pattern (Fig.~\ref{fig:1}c).

\section*{State-dependent localization}
\begin{figure}[!t]
    \centering
    \includegraphics[keepaspectratio,width=9cm]{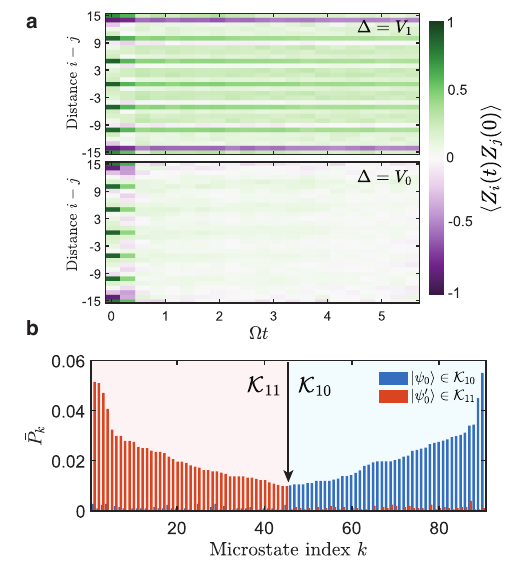}
        \caption{\textbf{Experimental studies of ergodicity-breaking.} \textbf{a}, Site-resolved spin autocorrelators $\langle Z_i(t)Z_j(0)\rangle$ for (top) the strongly fragmented $H_\text{LGT}$ and (bottom) the ETH-obeying case where $\Delta = V_0$, starting from a $\mathbb{Z}_5$-ordered state. Under $\Delta=V_0$, $\langle Z_i(t)Z_j(0)\rangle$ rapidly relaxes to zero everywhere, indicating the loss of initial state information. In $H_\text{LGT}$, the green streaks every 5 rows indicate that memory of the $\mathbb{Z}_5$-ordering quickly saturates and persists. \textbf{b}, Time-averaged $Z$-microstate projections for dynamics initialized from $\ket{\psi_0}$ (blue) and $\ket{\psi_0'}$ (orange), belonging to two distinct Krylov fragments $\mathcal{K}_{10}$ and $\mathcal{K}_{11}$ respectively (Table~\ref{table:1}). All microstates in the respective fragment are sampled, with leakages between fragments relatively smaller in magnitude. This relative suppression is highlighted by the step-like feature (arrow), beyond which bars represent product states in the other fragment. This feature shows that the non-ergodicity of autocorrelators in (a) is mainly due to fragmentation rather than disorder localization.
}
    \label{fig:2}
\end{figure}

To probe the dynamics of our system, we perform quenches starting from a $\mathbb{Z}_5$-ordered state (i.e., Rydberg atoms at every fifth site) and measure the site-resolved $Z$ autocorrelators  (Fig.~\ref{fig:2}a). Under the LGT mapping, the initial state transforms into a chain with $N_{c} = 5$ electric charge clusters. The dynamics achieved by setting $\Delta = V_{1}$ (Eq.~(\ref{eqn:LGTHamiltonian})) appear to retain memory of the initial state, as seen from the green streaks every five rows that initially decay but quickly saturate to finite values. The persistence of $\mathbb{Z}_{5}$ ordering reflects the presence of non-ergodicity, which is in stark contrast to the dynamics achieved under $\Delta = V_{0}$, where autocorrelators rapidly decay at every range along the chain to give a featureless plot. In the latter case, this corresponds to the PXQ + QXP model \cite{marcuzzi2017facilitation, magoni2021emergent}, which does not manifest any fragmentation.

The observed non-ergodicity is attributed to Hilbert space fragmentation, which we illustrate through the formation of disjointed subspaces. Under $H_\text{LGT}$, we quench initial states from two distinct Krylov fragments $\mathcal{K}_{10}$ and $\mathcal{K}_{11}$ (Table~\ref{table:1}) 
. Within each fragment, we then measure the time-averaged $Z$-basis microstate projections $\bar{P}_k = \frac{1}{t_f-t_i}\int_{t_i}^{t_f}|\langle e_k|\psi (t)\rangle|^2 \,dt$, where $\ket{e_k}$ is a $Z$-basis product state and the averaging times used are $\Omega t_i = 0.56(1)$ and $\Omega t_f = 5.60(1)$ (Fig.~\ref{fig:2}b). Projections onto microstates in the same fragment as the initial state are \textit{all} significant and much greater than onto microstates in the other fragment. The step-like feature at the center of Fig.~\ref{fig:2}b (black arrow), beyond which basis states start to belong to the other fragment, offers evidence that the fragmentation of $H_\text{LGT}$ is the dominant reason for the non-ergodicity we observe, rather than localizing effects from atomic position disorder \cite{marcuzzi2017facilitation}.

\begin{figure}[!t]
    \centering
    \includegraphics[keepaspectratio,width=9cm]{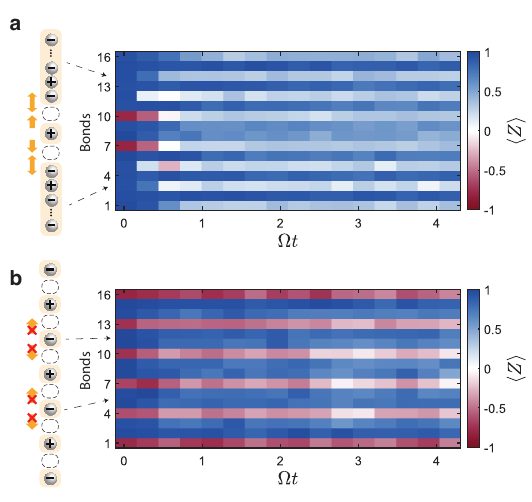}
        \caption{\textbf{State-dependent localization.} \textbf{a}, We start from two Rydberg atoms placed between three blocks of ground state atoms, where
        each block of ground-state atoms maps onto an electric charge cluster. The charge clusters are always spaced by at least two vacua (not depicted in detail here). Here we experimentally observe rapid spreading of the Rydberg excitations across the chain, which corresponds to a delocalization of the charge cluster boundaries. \textbf{b}, We start instead from a $\mathbb{Z}_3$-ordered state with Rydberg atoms at every third site, which corresponds to a saturated number of electric charge clusters. The state remains frozen under $H_\text{LGT}$. We attribute the observed minor delocalization mainly to state preparation errors and off-resonant scattering effects.
}
    \label{fig:3}
\end{figure}

Having verified the signatures of fragmentation, we now investigate a central feature of fragmented systems that motivates statistical locality. Distinct from many-body localized systems, localization behavior here depends highly on the initial state. We prepare in our experiment two different initial states. The first state consists of all atoms in the ground state, except for two Rydberg atoms near the center. Under the LGT mapping, this state consists of three charge clusters, where the two boundary charge clusters each comprise seven charged particles, leaving much room for these clusters to shrink and for the middle cluster to grow (Fig.~\ref{fig:3}a). In contrast, the second initial state is a $\mathbb{Z}_{3}$-ordered state with Rydberg atoms at every third site, which maps onto a saturated number of charge clusters, without any room for the clusters to expand or contract (Fig.~\ref{fig:3}b). Indeed, from single-atom resolved $Z$-measurements, we observe the spreading of Rydberg excitations over the chain for the first state, whereas the second state remains mostly localized.

\begin{figure*}[!htbp]
    \centering
    \includegraphics[keepaspectratio,width=18cm]{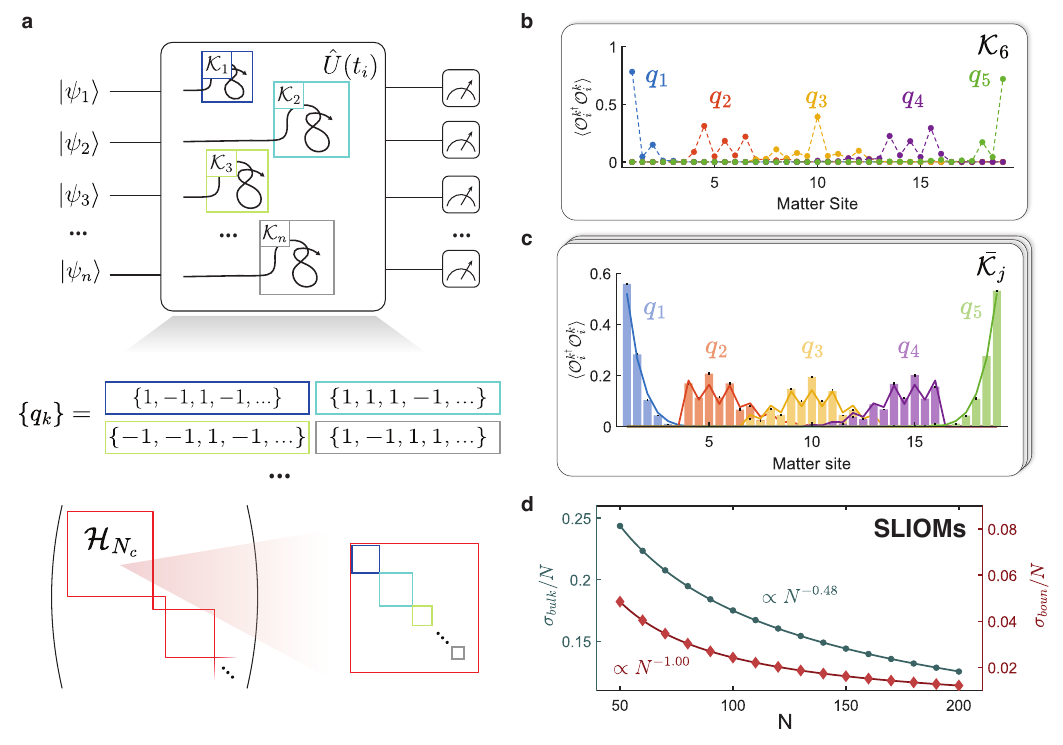}
        \caption{\textbf{Experimental observation of statistical localization.} \textbf{a}, 
        Statistical locality is examined through the spatial distribution of electric charge cluster densities for a typical state within a symmetry sector ($\mathcal{H}_{N_{c}}$). To observe statistical locality, we take as inputs the product states from each subordinate Krylov fragment, which is characterized by a unique set of eigenvalues of every conserved quantity $\{q_{k}\}$. We then time-evolve them for varying durations before applying a global projection measurement. Our measurements are averaged, weighted by the subspace dimension, to yield the typical-state spatial distributions. \textbf{b}, Experimentally measured spatial distributions of charge-cluster densities for time-averaged dynamics within a single Krylov fragment $\mathcal{K}_6$. The time-averaging is performed from $\Omega t_i = 0.56(1)$ to $\Omega t_f = 5.60(1)$ over 19 time steps. \textbf{c}, Spatial distribution of charge-cluster densities within the $N_c = 5$ symmetry sector. Our experimental reconstruction of a typical state (bars) shows good agreement with the theory prediction for an infinite-temperature ensemble of states (solid lines). All cluster densities appear partially localized to a sub-extensive region. \textbf{d}, Theoretical scaling behavior $N^{-\alpha}$ of the spatial extent of clusters in the bulk (blue) and at the boundary (red) of the chain. Clusters in the bulk appear to be partially localized in the thermodynamic limit ($\alpha = 0.48$), with their relative spatial extent $\sigma/N$ weakly vanishing for larger systems despite their absolute extent increasing. In contrast, the scaling exponent $\alpha = 1.00$ for the boundary clusters implies their complete localization in the thermodynamic limit.}
    \label{fig:4}
\end{figure*}

We note that state-dependent localization is not unique to strongly fragmented systems; it occurs also in weakly fragmented systems. However, the small number of localizing states in the latter is insufficient to impede thermalization for a \textit{typical} state in Hilbert space. In contrast, under strong fragmentation, the size of the largest connected sector in $H_\text{LGT}$ becomes vanishingly small, suggesting that almost all states fail to thermalize and instead localize. To verify this and quantify the degree of localizing behavior precisely, it is necessary to construct the conserved quantities of $H_\text{LGT}$. 

\section*{Statistically Localized Integrals of Motion}
Conserved quantities in a many-body system generally represent restrictions on information propagation\cite{hart2023exact,rademaker2017many,hahn2021information}. In the LGT model, the nonlocal conserved quantities $\mathfrak{q}_{k}$ identify the net charge of the $k^{\text{th}}$ charge cluster from the left of the chain:
\begin{equation}
    \mathfrak{q}_k = \sum_{i\in\{1,\frac{3}{2},2\cdots\}}^{N} (-1)^{2i} \,\mathcal{O}_{i}^k\, ,
\end{equation}
where $\mathcal{O}_{i}^k = \sum_{j=i\text{ mod }1}^{\text{min}(i-1,N-i)} P_{i-j}^{k,L}P^{k,R}_{i+j}$  denotes the center-of-mass projector of the $k^{\text{th}}$ charge cluster and $P^{k,L}_{i-j}$ ($P^{k,R}_{i+j}$) is the projector onto configurations where the leftmost (rightmost) electric charge on the $k^{\text{th}}$ cluster resides on site $i-j$ ($i+j$). The center of a cluster, denoted by the center-of-mass index $i$, may take on integer or half-integer values corresponding respectively to the cases where a cluster spans an odd or even number of sites (charged or neutral). The eigenvalues of $\mathfrak{q}_k$ are $1(-1)$ if the $k^{\text{th}}$ charge cluster is charged(neutral) and $0$ if the configuration has fewer than $k$ charge clusters. These operators $\mathfrak{q}_k$ are so-called ``statistically localized integrals of motion" (SLIOMs) \cite{rakovszky2020statistical}. Collectively, eigenvalues of the SLIOMs $\{q_k\}$ define \textit{every} Krylov fragment in $H_{\text{LGT}}$. For example, the initial state evolved in Fig.~\ref{fig:3}a corresponds to the sector $\{q_1 = 1,\: q_2=1, \: q_3=1, \: q_{4},q_5\cdots=0\}$. The projectors $P_{i}^{k,L},P_{i}^{k,R}$ are generally highly nonlocal operators but are diagonal in the $Z$-basis. Since our experiment measures the entire $Z$-bitstring in each snapshot, we can extract the nonlocal observables $\langle\mathcal{O}^{k\dag}_i\mathcal{O}^k_i\rangle$.

With the ability to measure the distributions of the conserved electric charge clusters in our system, we ask the following question: are these charge clusters statistically localized? In other words, are the spatial distributions of the corresponding SLIOMs localized for almost all states in Hilbert space? To investigate this, we employ the following procedure for sampling states within an $N_{c}$ symmetry sector $\mathcal{H}_{N_{c}}$ (Fig.~\ref{fig:4}a): we first initialize and time-evolve states belonging to every Krylov fragment in $\mathcal{H}_{N_{c}}$. We then sample states from a temporal ensemble by making $Z$-measurements at various times between $t_i$ and $t_f$ \cite{choi2023preparing, pilatowsky2023complete, pilatowsky2024hilbert, ghosh2024late}. Subsequently, averaging over all the sampled states within a given Krylov fragment corresponding to some net-charge pattern, we construct the spatial distribution of site-resolved charge cluster densities $\langle\mathcal{O}_i^{k\dag}\mathcal{O}_  i^k\rangle$ for that fragment. We note (e.g.\ from Fig.~\ref{fig:2}b) that all the $Z$-basis states within a given Krylov subspace have significant contribution, therefore the above sampling protocol is ergodic. Finally, we average these individual distributions, weighted by their fragment sizes, to ensure that we sample from an appropriate measure over $\mathcal{H}_{N_{c}}$.

In our experiment, we probe the symmetry sector $N_{c} = 5$, sampling at least 3800 states from each Krylov fragment. We additionally employ a post-selection scheme to project measured states into $\mathcal{H}_{N_c}$. Fig.~\ref{fig:4}b shows the average distribution for states within a single Krylov fragment given by the SLIOMs $\{q_k\}=\{1,-1,1,-1,1\}$. We start to see that the distributions of SLIOMs across the chain remain confined to a localized section. We also see a jagged feature in all the distributions. This tells us that the SLIOMs remain well-conserved experimentally; for instance, the distribution of the second cluster from the left only has significant contributions from half-integer sites, which is expected from a neutral cluster ($q_2 = -1$). 

Appropriately patching together these distributions from other fragments, we construct a sample average of SLIOM distributions over the entire symmetry sector (Fig.~\ref{fig:4}c). The distributions appear smoother and still remain partially localized, verifying the statistical locality of the clusters. These distributions have unequal contributions from half-integer sites and integer sites. This comes from the fact that charged clusters tend to occupy, on average, fewer sites than neutral clusters since their minimal length is one (instead of two). In turn, the projector $O^k_i$ for charged clusters ($i\in\mathbb{Z}$) projects onto a larger Hilbert space. Our experimentally sampled distributions resemble that for a typical state, showing good agreement with predictions for the infinite-temperature state projected within $\mathcal{H}_{N_{c}}$.

From Fig.~\ref{fig:4}c, two questions immediately arise: is this statistical locality an artifact of small system sizes and does it remain when sampling states randomly over the entire Hilbert space? To investigate this, we numerically compute the infinite-temperature distributions of SLIOMs $q_k$ and extract their widths $\sigma$ (full-width half maxima) for system sizes ranging from $N=50$ to $N=200$ (Fig.~\ref{fig:4}d). Here we average over all the states hosted in the subspace with no nearest-neighbor Rydberg atoms. We find that SLIOM widths in the bulk, normalized to the system size, scale approximately as $N^{-1/2}$. The scaling exponent means that for almost all states, information in the system only propagates across a subsection of the chain whose extent scales as $N^{1/2}$. Although the spatial extents of charge clusters increase with the system size, they only do so weakly, occupying a vanishing fraction of the chain for large systems (i.e., partially localized). On the other hand, the normalized width of boundary charge clusters is observed to scale almost exactly as $N^{-1}$, indicating that these clusters remain completely localized to a finite extent in space, even for large systems. 

We have demonstrated the first experimental signatures of statistical localization on a Rydberg-atom lattice gauge theory quantum simulator with strong fragmentation. The dynamics of localized charge clusters are interesting for physics at both low and high energies. From a high-energy perspective, simulations of charge cluster dynamics may be important for understanding the non-perturbative unfolding of scattering events in particle colliders \cite{Bauer:2023qgm,le2023observation}. While the charge clusters remain localized and stationary in this study, a possible extension would be to explore conditions that can give the charge clusters a finite momentum kick in one dimension, while keeping the cluster width finite. At the low energy scale, the statistical localization at the edges is reminiscent of strong zero modes, where information is protected at the boundaries of a system for nearly infinite time, even at infinite temperatures \cite{fendley2016strong,olund2023boundary,else2017prethermal}. This is in contrast to regular edge modes in topological systems, which are protected by a finite energy gap \cite{kempkes2019robust,verresen2018topology}. In particular, boundary SLIOMs in the strongly fragmented $t-J_z$ model predict a gapless ground state that is described by a topologically non-trivial symmetry-enriched critical point \cite{rakovszky2020statistical,verresen2018topology,verresen2021gapless}. 
Our strongly fragmented but distinct model may also yield unresolved topological structures, prompting further study. Further, our measurements of statistical locality, complementary to the subsystem Loschmidt echo scheme \cite{karch2025probing}, may be used to distinguish between strongly and weakly fragmented systems and subsequently observe the recently conjectured class of freezing phase transitions \cite{morningstar2020kinetically, wang2023freezing}. 
Finally, the survival of strong fragmentation under $H_\text{LGT}$ in higher dimensions remains an open question for future investigation.


\clearpage
\renewcommand{\figurename}{Extended Data Fig.}
\setcounter{figure}{0}

\section*{Methods}
\section*{Experiment Sequence}

Our experiments start from a cooled atomic cloud of $^{87}$Rb atoms. We load single atoms into a two-dimensional array of 80 static optical tweezers with $1.0$~mK depth, realizing a loading rate of $\geq 80\%$ using $D_1$ $\Lambda$-enhanced loading. The atoms are subsequently rearranged using mobile tweezers to form three defect-free chains with $N_{a} = 16$ atoms, achieving a success probability of about $70\%$ with a multitweezer algorithm \cite{tian2023parallel}. We then perform two additional stages of cooling: $D_{1}$ $\Lambda$-enhanced gray molasses cool the atoms down to $37(2)~\mu$K, followed by a preliminary implementation of Raman sideband cooling in 0.6~mK traps for 8 ms, resulting in a final atom temperature of $17(1)~\mu$K (determined from Monte Carlo fits). We optically pump the atoms to  initialize them in the ground state $\ket{g} = \ket{5S_{1/2}, F = 2, m_F = 2}$. At this point, the optical tweezers are turned off, following which we apply single-site addressing light to a subset of array sites along with a global resonant Rydberg $\pi$-pulse to prepare product states of ground and Rydberg atoms \cite{labuhn2014single}. A second global Rydberg pulse is applied for a variable duration and detuning to probe the dynamics of interest. A $3~\mu$s microwave pulse is applied to ionize Rydberg atoms. Finally, imaging beams are applied to read out the ground (Rydberg) state as the presence (absence) of an atom. The combined state preparation and measurement fidelities for the ground and Rydberg states are 99\% and 95\% respectively. 

Our Rydberg scheme excites atoms to the Rydberg state $\ket{r} = \ket{70S_{1/2}, m_J = 1/2}$ using a two-photon excitation with counterpropagating $\sigma^+$-polarized 420~nm light and $\sigma^-$-polarized 1013~nm light. The 420~nm light is red detuned by $\Delta' \approx 2\pi \times 780$ MHz from the intermediate state $\ket{e} = \ket{6P_{3/2}, F = 3, m_{F} = 3}$ and the two-photon Rydberg Rabi frequency is $\Omega = 2\pi \times 1.39(1)$~MHz. We measure the damping time for ground-Rydberg Rabi oscillations to be $\tau = 27(6)~\mu$s for a single atom, which is about an order of magnitude longer than the timescale of dynamics probed in this work. The three defect-free chains prepared after rearrangement are spaced by about $3R_b$ such that inter-chain Rydberg interactions are on the scale of kHz, negligible in comparison to the Rabi frequency. 

For the experiments on $H_\text{LGT}$, the interatomic spacing is set to $a = 3.37~\mu$m, resulting in a measured interaction strength of $V_1 =2\pi\times 9.2(1)$~MHz. This spacing is doubled instead to 6.74~$\mu$m for experiments on $H_{V_0}$, thereby realizing $V_0 =2\pi\times 9.2(1)$~MHz. For preparation of the $\mathbb{Z}_3$ ordered state, during single-site addressing we additionally impart a small positive detuning of about $\Delta \approx 2\pi \times 300$~kHz to the global Rydberg $\pi$-pulse to compensate for longer-range interactions on the order of the next-next-nearest-neighbor interaction $V_2$. Each data point shown in this manuscript is averaged from at least 200 defect-free chain samples. 

\section*{Effects of position disorder}
\label{app:pos_disorder}
\begin{figure}[!htbp]
    \centering
    \includegraphics[keepaspectratio,width=9cm]{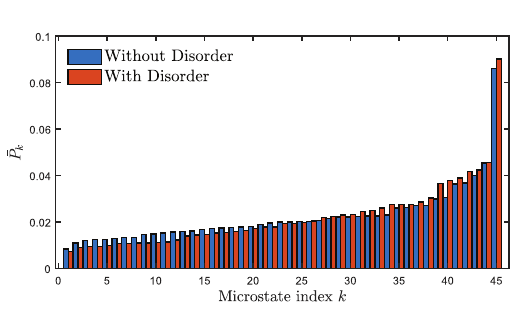}
        \caption{\textbf{A comparison of dynamics with (red bars) and without (blue bars) position disorder.} We plot time-averaged $Z$-microstate projections $\bar{P}_k$ evolving from an initial state within $\mathcal{K}_{11}$ (Table~\ref{table:1}). The two distributions appear similar, indicating only a minor localizing effect from the position disorder.
}
    \label{fig:posdisorder}
\end{figure}

We realize $H_\text{LGT}$ on our simulator by matching the detuning to the next-nearest-neighbor interaction: $\Delta = V_1$. However, finite atomic temperature results in position disorder, which leads to interaction fluctuations between the atoms. For $T=17(1)~\mu$K, we estimate a radial position disorder of $\sigma_r = 0.083~\mu$m and a resulting interaction fluctuation of about $\delta V_1 = 6|V_1|\sigma_r/(2a) = 2\pi \times 0.68$~MHz. Although this is less than our Rabi frequency $\Omega = 2\pi \times 1.39(1)$~MHz, it may still result in Anderson-like localization effects. If significant, such localization would be detrimental to our Krylov fragment sampling procedure described in the main text, hence it is necessary to evaluate its effects. We numerically simulate dynamics in the fragment $\mathcal{K}_{11}$ from the initial state reported in Table~\ref{table:1}, both with and without disorder. We then average the $Z$-basis microstate projections from $\Omega t_i' = 7$ to $\Omega t_f' = 14$ and plot $\Bar{P}_k$ for states in $\mathcal{K}_{11}$ to evaluate if the position disorder significantly affects this distribution (Extended Data Fig.~\ref{fig:posdisorder}). The difference between the largest and smallest state projection within the fragment is slightly greater when disorder corresponding to our experimental parameters is added, indicating a small localizing effect. However, the distributions appear largely similar, and we conclude that the effects of disorder are not significant enough to be detrimental to our sampling procedure.

\section*{Fragment Parameters}
\label{app:frag_params}

\begin{table*}[!htbp]
\centering
\begin{tabular}{ ||c|c|c|c|c|c|| } 
 \hline
 No.& Net-charge pattern & SLIOMs $\{q_k\}$ &$\{N_q,N_0\}$ &Fragment size & Initial atomic state\\
 \hline
$\mathcal{K}_1$ & \textbf{c c c c c} & $1,1,1,1,1$ & $5,0$ & 165 & $\ket{rggggrggggrggggr}$\\ 
   \hline
$\mathcal{K}_2$ & \textbf{c c c n n} & $1,1,1,-1,-1$ & $3,2$ & 45 & N/A\\ 
   \hline
$\mathcal{K}_3$ & \textbf{c c n c n} & $1,1,-1,1,-1$ & $3,2$ & 45 & N/A\\ 
   \hline
$\mathcal{K}_4$ & \textbf{c c n n c} & $1,1,-1,-1,1$ & $3,2$ & 45 & N/A\\ 
   \hline
$\mathcal{K}_5$ & \textbf{c n c c n} & $1,-1,1,1,-1$ & $3,2$ & 45 & N/A\\ 
   \hline
$\mathcal{K}_6$ & \textbf{c n c n c} & $1,-1,1,-1,1$ & $3,2$ & 45 & $\ket{rgggggrggrgggggr}$\\ 
   \hline
$\mathcal{K}_7$ & \textbf{c n n c c} & $1,-1,-1,1,1$ & $3,2$ & 45 & $\ket{rgggggrgggrggggr}$\\ 
   \hline
$\mathcal{K}_8$ & \textbf{n c c c n} & $-1,1,1,1,-1$ & $3,2$ & 45 & $\ket{grggggrggrggggrg}$\\ 
   \hline
$\mathcal{K}_9$ & \textbf{n c c n c} & $-1,1,1,-1,1$ & $3,2$ & 45 & $\ket{grggggrggggrgggr}$\\ 
   \hline
$\mathcal{K}_{10}$ & \textbf{n c n c c} & $-1,1,-1,1,1$ & $3,2$ & 45 & $\ket{grggggrgggrggggr}$\\ 
   \hline
$\mathcal{K}_{11}$ & \textbf{n n c c c} & $-1,-1,1,1,1$ & $3,2$ & 45 & $\ket{grgggrggggrggggr}$\\ 
   \hline
$\mathcal{K}_{12}$ & \textbf{c n n n n} & $1,-1,-1,-1,-1$ & $1,4$ & 9 & N/A\\ 
 \hline
$\mathcal{K}_{13}$ & \textbf{n c n n n} & $-1,1,-1,-1,-1$ & $1,4$ & 9 & N/A \\ 
  \hline
$\mathcal{K}_{14}$ & \textbf{n n c n n} & $-1,-1,1,-1,-1$ & $1,4$ & 9 & $\ket{grgggrggggrgggrg}$\\ 
  \hline
$\mathcal{K}_{15}$ & \textbf{n n n c n} & $-1,-1,-1,1,-1$ & $1,4$ & 9 & $\ket{grgggrgggggrggrg}$\\ 
 \hline
$\mathcal{K}_{16}$ & \textbf{n n n n c} & $-1,-1,-1,-1,1$ & $1,4$ & 9 & $\ket{grgggrgggggrgggr}$\\ 
 \hline
\end{tabular}

\caption{A consolidation of parameters for Krylov fragments within $\mathcal{H}_{N_c=5}$}
\label{table:1}
\end{table*}

\begin{figure}[!htbp]
    \centering
    \includegraphics[keepaspectratio,width=9cm]{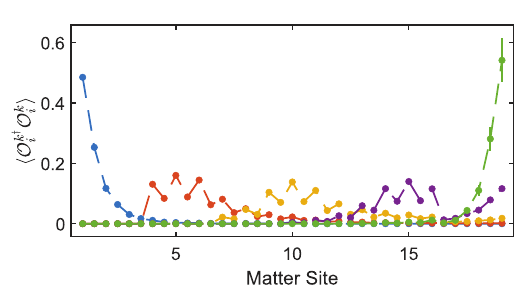}
        \caption{\textbf{SLIOM distributions sampled from $\mathcal{H}_{N_c=5}$, but without postselection into $\mathcal{S}$.} The distributions are more spread out towards the right side of the chain. This is because state preparation infidelity, our main source of imperfection, is more likely to result in initializing a state with $N_c'<5$ compared to $N_c'>5$ (since $\mathcal{F}_{g} > \mathcal{F}_{r}$). States belonging to a symmetry sector with $N_c<5$ would naturally have their conserved charges more spread out across the chain, resulting in the broadening of distributions.
}
    \label{fig:sliom_nopostsel}
\end{figure}

Here, we outline some parameters for the Krylov fragments probed in the $\mathcal{H}_{N_c=5}$ symmetry sector. The symmetry sector $\mathcal{H}_{N_c=5}$ has sixteen disconnected fragments within it. In our experiments, it was only necessary to probe ten of them. This is because fragments $2,3,4,5,12$, and $13$ (from Table~\ref{table:1}) have SLIOMs that are odd under inversion symmetry. We exploit this symmetry and simply pick out states that are the inversion symmetric partners of states sampled in these fragments, to sample from the remaining six fragments. For each of the fragments, eigenvalues of the first five SLIOMs are listed, with all other eigenvalues $q_{6}, q_{7}, \cdots=0$. Initial states are chosen from each of these subspaces with a minimal number of Rydberg excitations since the dominant experimental imperfection is the Rydberg state preparation infidelity.

\section*{Data Processing and Numerical Details}

Our raw experimental data comprises single-cycle $Z$-basis bitstrings. We do a first round of post-selection to discard any bitstrings that violate nearest-neighbor blockade (with two or more consecutive Rydberg atoms). For the data used in Fig.~\ref{fig:4}b,c to sample states in the $N_c=5$ symmetry sector, we further discard states that violate $N_c$ conservation. The bitstrings are then mapped to determine cluster lengths and positions. To measure the SLIOMs $\{q_k\}$, each bitstring is processed sequentially from left to right, where we track the center-of-mass positions and ordering of all the cluster parities. Extended Data Fig.~\ref{fig:sliom_nopostsel} shows the SLIOM distributions, as in Fig.~\ref{fig:4}c, without $N_c$ postselection. The infinite-temperature SLIOM distribution in Fig.~\ref{fig:4}c was obtained by enumerating the cluster positions for each $Z$-basis microstate in the $N_c=5$ sector and uniformly averaging them. 

To obtain the global infinite-temperature distributions for larger system sizes used in Fig.~\ref{fig:4}d, we evaluate the analytical expressions Eq.~(\ref{boundary_analytical}) and Eq.~(\ref{bulk_analytical}). Cluster widths $\sigma$ for each SLIOM distribution are obtained by fitting cluster distributions to a Gaussian profile and equating the width as the full-width half maximum of the fit, i.e.\ $\sigma = 2\sqrt{2\ln{2}}\sigma_{0}$, where $\sigma_{0}^2$ is the variance of the Gaussian fit. The widths $\sigma$ plotted in Fig.~\ref{fig:4}d are the result of averaging the widths of all SLIOM distributions in the bulk, weighted according to their operator support. Since boundary SLIOMs are excluded, the symmetry sectors $N_c=1,2$ are excluded as well. For the larger system sizes that we consider in our numerics, these symmetry sectors occupy a negligible fraction of Hilbert space. For the distribution $q_k$, this is equal to the fraction of states in Hilbert space that contain at least $k$ clusters.

\section*{Effective Hamiltonian}
\label{app:deviation}
The Hamiltonian of our open Rydberg-atom chain can be written as:
\begin{equation} \label{eq:fullH}
    H_{\text{Ryd}} = -\Delta \sum_{i=1}^{N_{a}} Q_{i} + \frac{\Omega}{2}\sum_{i=1}^{N_{a}} X_{i} + \sum_{j = 1}^{3} \sum_{i = 1}^{N_{a}-j} V_{j-1} Q_{i} Q_{i+j} \, .
\end{equation}
Note that for an open chain, we can treat boundary terms identically to bulk terms if we place two virtual-ground state atoms next to each boundary of the chain (i.e., \emph{$g$-padding})\cite{yang2025probing}. We henceforth work with an effective chain of length $N = N_{a} + 4$.

We first neglect van-der-Waals interactions of range four and longer, since they result in energy offsets of at most $V_3 = 2\pi\times  2.3$~kHz, orders of magnitude smaller than $\Omega = 2\pi\times  1.39(1)$~MHz. Then, there are four important energy scales in $H_\text{LGT}$: $V_0 \gg V_1 = \Delta \gg \Omega \gg V_2$. The large interaction energies $V_1, V_0$ motivate splitting the Hamiltonian $H_{\text{Ryd}} = H_0 + H_1 + H_{\text{pert}} + H_{\text{LR}}$:
\begin{widetext}
\begin{equation}
    H_0 = V_0\sum_{i=1}^{N-1}Q_{i}Q_{i+1} \quad , \quad
    H_1 = -\Delta \sum_{i=1}^{N} Q_{i} + V_1\sum_{i=1}^{N-2}Q_{i}Q_{i+2} = -V_1\sum_{i=1}^{N-2}Q_{i}P_{i+2}  \quad , \quad
    H_{\text{pert}} = \frac{\Omega}{2}\sum_{i=3}^{N-2} X_{i}
\label{eq:H01pert}
\end{equation}
and $ H_{\text{LR}} = V_2\sum_{i=1}^{N-3}Q_{i}Q_{i+3}$, the longer-range van-der-Waals coupling. The second equality for $H_1$ is obtained by setting $\Delta = V_1$. 

Note that from Eq.~(\ref{eq:H01pert}) onwards, we have added the fictitious terms $-\Delta \sum_{j\in\{1,2,N-1,N\}}Q_j + V_0\sum_{j\in\{1,2,N-2,N-1\}}Q_jQ_{j+1} + V_1\sum_{j\in\{1,2,N-3,N-2\}}Q_jQ_{j+2}=0$ in order to massage the Hamiltonian. In the new Hilbert space under \emph{$g$-padding}, these fictitious terms always evaluate to zero, hence the addition is valid. $H_0, H_1$ are the dominant terms, and block diagonal in the $Z$-basis, with blocks separated by energies $V_0$ and $V_1$ respectively. 

We treat $H_{\text{pert}}$ as a perturbation that can couple states in distinct blocks given by $V_0$ and $V_1$. First focusing on the largest energy scale $V_0$, blocks are defined by the number of nearest-neighbor Rydberg pairs $n_{\text{NN}}$. Entering the rotating frame with respect to $V_0$, we have \cite{zhao2025observation}:
\begin{equation} \label{fourier}
    H' = \frac{\Omega}{2}\sum_{i=2}^{N-1}\tilde{\sigma}^{+}_{i}\left[\left(P_{i-1}P_{i+1}\right) + e^{iV_{0}t}\left(Q_{i-1}P_{i+1}+P_{i-1}Q_{i+1}\right) + e^{2iV_{0}t}\left(Q_{i-1}Q_{i+1}\right)\right] + H_1 + H_{\text{LR}} + \text{H.c.} \, ,
\end{equation}
where $\tilde{\sigma}^{-} = \ket{r}\bra{g}$ and $\tilde{\sigma}^{+} = \ket{g}\bra{r}$. We hence identify the Fourier components of $H_{\text{pert}}$ in this frame as:
\begin{equation}
    T_0 = \frac{\Omega}{2}\sum_{i=2}^{N-1}P_{i-1}X_{i}P_{i+1} \quad , \quad
    T_{\pm 1} = \frac{\Omega}{2}\sum_{i=2}^{N-1} \tilde{\sigma}_{i}^{\pm}\left(Q_{i-1}P_{i+1}+P_{i-1}Q_{i+1}\right) \quad , \quad
    T_{\pm 2} = \frac{\Omega}{2}\sum_{i=2}^{N-1}Q_{i-1}\tilde{\sigma}_{i}^{\pm}Q_{i+1}
\end{equation}
Physically, these are generalized ladder operators that increase or decrease $n_{\text{NN}}$ by zero, one or two. They correspondingly obey the commutation relations $[H_0,T_m] = mV_0T_m$, coupling states across an energy scale $mV_0$. With this structure, we construct the effective Hamiltonian by performing a Schrieffer-Wolff transformation, expanding in orders of the small parameter $\frac{\Omega}{V_0}$ \cite{bravyi2011schrieffer}. The leading order Hamiltonian is \cite{bluvstein2021controlling}:
\begin{equation}
    H_P = H_0 + H_1 + T_0 + H_{\text{LR}} = n_{\text{NN}} -V_1\sum_{i=1}^{N-2}Q_{i}P_{i+2} + \frac{\Omega}{2}\sum_{i=2}^{N-1}P_{i-1}X_{i}P_{i+1} + H_{\text{LR}} \, .
\end{equation}

In all our experiments, nearest-neighbor blockade dominates and $n_{\text{NN}}$ is well conserved. We hereafter set $n_{\text{NN}} = 0$, corresponding to the nearest-neighbor blockaded subspace. Higher-order terms consist of self-energies and correlated spin-flips mediated by states lying outside the nearest-neighbor blockaded subspace. The second-order terms have amplitudes $\frac{\Omega^2}{4V_0} \approx 2\pi \times 0.83$~kHz, several orders of magnitude smaller than $V_0, V_1, \Omega$ and hence expected to be negligible. 

We now focus on the next dominant energy scale, $V_1$. This time, we split $H_P$ into the dominant block-diagonal term $H_1$ and the perturbation $T_0$. Blocks in $H_1$ are defined by the quantity $N_c = 1 + \sum_{i=1}^{N-2}Q_{i}P_{i+2} = 1 + \sum_{i=2}^{N-2}P_{i-1}Q_{i}P_{i+1}P_{i+2}$, where $N_c$ corresponds to the cluster number referenced in the main text. Similarly to above, we further split $T_0$ into generalized ladder operators $T_0 = T_{0,-1} + T_{0,0} + T_{0,1}$:

\begin{align}
    T_{0,-1} = & \frac{\Omega}{2}\sum_{i=3}^{N-2}P_{i-2}P_{i-1}\tilde{\sigma}^-_{i}P_{i+1}P_{i+2} + Q_{i-2}P_{i-1}\tilde{\sigma}^+_{i}P_{i+1}Q_{i+2} \\
    T_{0,0} = & \frac{\Omega}{2}\sum_{i=3}^{N-2}(P_{i-1}X_{i}P_{i+1})(P_{i-2}Q_{i+2} + Q_{i-2}P_{i+2}) \\
    T_{0,1} = & \frac{\Omega}{2}\sum_{i=3}^{N-2}P_{i-2}P_{i-1}\tilde{\sigma}^+_{i}P_{i+1}P_{i+2} + Q_{i-2}P_{i-1}\tilde{\sigma}^-_{i}P_{i+1}Q_{i+2} 
\end{align}
The ladder operators $T_{0,m}$ correspond to terms in $T_0$ that couple across an energy scale $mV_1$ and are accordingly defined by the commutators $[H_1,T_m] = mV_1T_m$. 

We now construct an effective Hamiltonian within a $N_c$ block by performing a Schrieffer-Wolff transformation, expanding in orders of the small parameter $\frac{\Omega}{V_1}$. We find the leading order Hamiltonian as:
\begin{equation}
    H^{(1)}_{\text{eff}} = H_1 + T_0 + H_{\text{LR}}= - V_1\sum_{i = 2}^{N-2}  P_{i-1}Q_{i}P_{i+1}P_{i+2} + \frac{\Omega}{2}\sum_{i=3}^{N-2}(P_{i-1}X_{i}P_{i+1})(P_{i-2}Q_{i+2} + Q_{i-2}P_{i+2}) + H_{\text{LR}}
\end{equation}
Projecting within a $N_c$ block, the first term is a constant determined by $N_c$ and is hence a constant. The third term consists of range-three van-der-Waals couplings of amplitude $V_2 \approx 2\pi\times 140$~kHz, relatively small compared to $\Omega$. Lastly, the second term, which we refer to as the ``PPXPQ + QPXPP'' model, corresponds to $H_\text{LGT}$ that we consider in the main text. 

The second-order effective Hamiltonian consists of self-energies and correlated spin-flips mediated by states outside the $N_c$ block. In terms of the ladder operators, it is equal to $\frac{[T_{0,1},T_{0,-1}]}{V_1}$. Explicitly, we have:
\begin{equation}
    H^{(2)}_{\text{eff}} = \frac{\Omega^2}{4V_1} \left[\sum_{i = 3}^{N-2}P_{i-1}Z_{i}P_{i+1}\left(P_{i-2}P_{i+2} - Q_{i-2}Q_{i+2}\right) + \sum_{i = 4}^{N-3} P_{i-2}\tilde{\sigma}^-_{i-1}P_{i}\tilde{\sigma}^+_{i+1}P_{i+2}\left(Q_{i-3}Q_{i+3}+P_{i-3}P_{i+3}\right) + \text{H.c.}\right]
    \label{eq:secondorder}
\end{equation}
\end{widetext}
The first term is a neighbor-dependent diagonal energy shift while the second term corresponds to constrained next-nearest-neighbor spin exchange. For our experimental parameters, these second-order couplings are of magnitude $\frac{\Omega^2}{4V_1} \approx 2\pi \times 53$~kHz and relatively small for the timescales of a few $\mu$s that we probe here. In fact, both these terms in Eq.~(\ref{eq:secondorder}) preserve the fragmentation in $H_\text{LGT}$. 

\begin{figure*}[!htbp]
    \centering    \includegraphics[keepaspectratio,width=18cm]{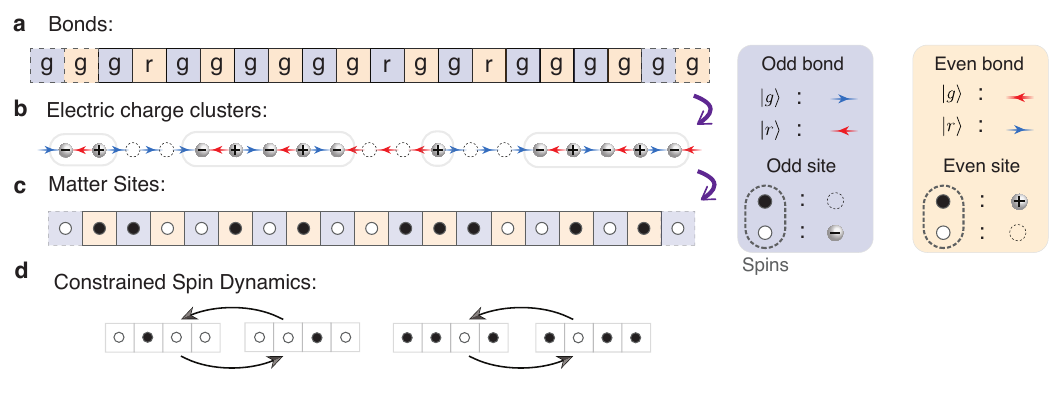}
        \caption{\textbf{Mapping from a chain of Rydberg atoms to a $U(1)$ LGT.} \textbf{a,} The atomic chain of ground and Rydberg atoms is first mapped to \textbf{b,} a chain of electric field strings on the bonds and charge clusters on the sites. Internal states of the atoms (ground or Rydberg) are mapped directly to electric field variables on the bonds, where right-facing strings alternate in correspondence between ground and Rydberg state atoms on odd and even bonds (vice versa for left-facing strings). \textbf{c,} Another mapping is performed to finally arrive at a spin representation. On odd sites, spin down (spin up) corresponds to a negative charge (vacua), whereas on even sites, spin down (spin up) corresponds to a vacua (positive charge). \textbf{d,} Tracing out the bond degrees of freedom leads to constrained spin dynamics on the matter sites, where spin-exchange is allowed between adjacent sites, only if the surrounding two sites are both spin-up or spin-down.} 
    \label{fig:ext_fig_mapping}
\end{figure*}

\section*{Mapping to \texorpdfstring{$H_\text{LGT}$}{H(LGT)}}
\label{app:particle_map}

We detail here the mapping from the ``PPXPQ + QPXPP'' Hamiltonian on a one-dimensional lattice of $N_a+4$ two-level atomic systems, to $H_\text{LGT}$ on a complementary lattice of $N_a+3$ spins characterizing a quantum link model \cite{surace2020lattice}.  
To do so, we begin by interpreting the atomic two-level configurations constrained by the Rydberg blockade as a staggered representation of a $\mathbb{Z}_2$ electric flux on lattice bonds~(Extended Data Fig.~\ref{fig:ext_fig_mapping}a,b).
Charged matter on sites of this lattice can then be assigned uniquely by Gauss's law.
If both electric strings point towards (away) from a site, a negative (positive) electric charge is assigned to be present at the site; if both strings point in the same direction, the site is assigned to be an unoccupied vacuum instead. 
More explicitly, with bond operators expressed in this electric basis as, 
\begin{align}
    & S^{-}_{(j,j+1)} = |\rightarrow\rangle\langle \leftarrow | \ , \\
    & S^{+}_{(j,j+1)} = |\leftarrow\rangle\langle \rightarrow | \ , \\
    & S^z_{(j,j+1)} =  \frac{|\leftarrow\rangle\langle \leftarrow | - |\rightarrow \rangle\langle \rightarrow|}{2} \ ,
\end{align}
and site electric charge densities of the form ${\rho^{z}_{j}=-\ket{-}\bra{-}_j}$ on odd sites and  $\rho^{z}_{j} = \ket{+}\bra{+}_j$ on even sites, the $U(1)$ gauge symmetries are:
\begin{equation}
    G_j = S^z_{(j-1,j)} - S^z_{(j,j+1)} - \rho^z_j \ \ \ .
\end{equation}
In the present staggered mapping to electric flux, the nearest-neighbor blockaded subspace can thus be expressed as a gauge-invariant subspace, $G_j\ket{\psi}=0$ for all sites $j$, for a system in which the matter is also staggered, i.e., positive (negative) charges are not allowed on odd (even) sites. 
From this perspective, the nearest-neighbor Rydberg blockade constraint thus endows a $U(1)$ local symmetry at each site of the lattice.
This picture of electric charges and strings will be our main frame of reference for visualizing statistical localization. 

In terms of the electric charges on the sites,  the five-link \enquote{PPXPQ + QPXPP} Hamiltonian expressing an active central atom translates to a four-site Hamiltonian with an active center pair of electric charges.  
Note that interactions remain local even when the gauge field is integrated out in this manner, indicating that local kinetic constraints still exist to govern the dynamics of the electric charge and vacuum clusters. 

Furthermore, the sequential pair of ground-ground (Rydberg) projectors \enquote{$PP$} ( \enquote{$PQ$}) translates to projectors onto a charged (vacuum) site. Hence, the local kinetic constraint on the charges is such that pair production occurs only at the boundaries of a pre-existing charge cluster, while retaining regions of vacuum that buffer the clusters.

In particular, the dynamics obey the following rules (Fig.~\ref{fig:1}b):
\begin{enumerate}
\item Clusters of electric charge can expand or contract by pairs comprising a positive charge and a negative charge, but cannot merge; the total number of clusters remains conserved.
\item Vacuum clusters must contain an even number of sites. 
\item As a result of $g$-padding, one end of an electric charge cluster is fixed to each lattice boundary.
\end{enumerate}
The above constrained dynamics leads to an invariant ordering of net charges across the lattice throughout the evolution.

As a convenient final step, we map the electric charge on lattice sites to spin-1/2 degrees of freedom~(Extended Data Fig.~\ref{fig:ext_fig_mapping}c).
In terms of these spin degrees of freedom, the Hamiltonian can be written with translationally invariant spin operators, i.e., all staggering can be accounted for with this final mapping.
The four-site Hamiltonian expressing pair creation at a charge-cluster boundary discussed above simply becomes a hopping term controlled on equal spin states of next-nearest neighbors,
\begin{equation}
    H_{\text{LGT}} = -w\sum_{j=3}^{N-2}\Pi_{j-1,j+2} \sigma^+_j\mathbin{S^-_{(j,j+1)}}\sigma^-_{j+1} + \text{H.c.}
\end{equation}
where $\Pi_{j-1,j+2} = \ket{\circ_{j-1}\circ_{j+2}}\bra{\circ_{j-1}\circ_{j+2}} + \ket{\bullet_{j-1}\bullet_{j+2}}\bra{\bullet_{j-1}\bullet_{j+2}}$, $\sigma_j^+ = \ket{\bullet_{j}}\bra{\circ_j}$, and $\sigma_j^- = \ket{\circ_j}\bra{\bullet_j}$. 
We note that without the link operator $S^{-}_{(j,j+1)}$, the Hamiltonian would be reminiscent of the first term in the models studied in References~\cite{yang2020hilbert, yang2025probing}.

\begin{figure*}[!htbp]
    \centering
    \includegraphics[keepaspectratio,width=18cm]{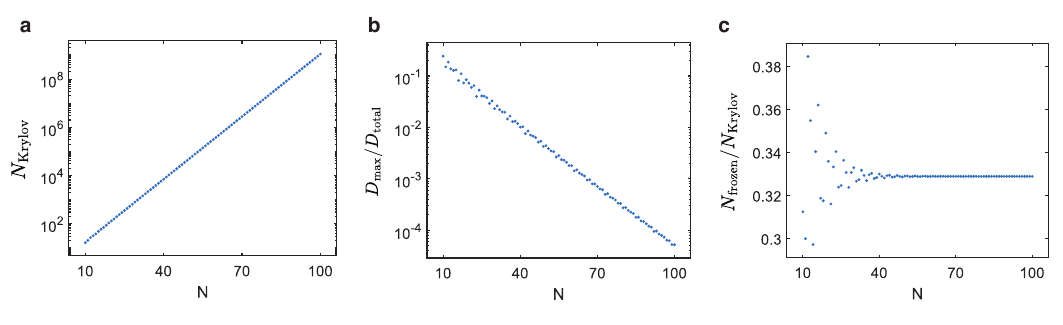}
        \caption{\textbf{Strong fragmentation in $H_\text{LGT}$.} \textbf{a}, The number of Krylov fragments, $N_{\text{Krylov}}$, scales exponentially as $1.22^N$. \textbf{b}, The size of the largest fragment $D_{\text{max}}$ as a fraction of the total Hilbert space dimension $D_{\text{total}}$ vanishes for large $N$, indicating strong fragmentation. \textbf{c}, The fraction of fragments that are frozen (dimension one) approaches a finite value close to $1/3$.
}
    \label{fig:figure6}
\end{figure*}

\section*{Hilbert space structure}
\label{app:fragmentation}
In this section, we examine the Hilbert space structure of $H_\text{LGT}$, mainly employing combinatorial arguments.
\subsection*{Counting of Krylov fragments within a symmetry sector}

First, we look at how many Krylov fragments we expect there to be within a given symmetry sector of $N_c$ charge clusters. Each Krylov sector is identified by a unique set of SLIOMs, or equivalently a net-charge pattern of electric charge clusters that remains conserved across the chain. Since the number of vacuum sites between electric charge clusters is even, the net charge of charged clusters also alternates over the chain. A consequence of mapping with $g$-padding and bond/site labels beginning with 1 is that the net-charge of the first charged charged cluster is always negative. Hence, the pattern of net-charges for charged clusters is pre-determined to be $\{-,+,-\cdots\}$. In other words, there are only two independent choices for each cluster - charged ``c'' or neutral ``n'', leading initially to $2^{N_c}$ valid patterns. Finally, the total number of sites imposes a parity constraint on the number of charged clusters in the chain, since only they occupy an odd number of sites. Thus, in a system with even (odd) sites, the number of charged clusters must also be even (odd). This reduces the total number of valid patterns by half. As a result, the number of valid patterns, and thus the number of Krylov fragments in $\mathcal{H}_{N_c}$, is given by $2^{N_c} / 2 = 2^{N_{c}-1}$.

\subsection*{Krylov subspace dimensions and strong fragmentation}
A generic product state consists of a variable number of charge clusters. Here, we combinatorially find exact expressions of the dimensions and number of Krylov fragments, relying on the fact that charged (neutral) clusters always occupy an odd (even) number of sites.

Suppose we have an $N_a$ atom chain. Under the LGT mapping, we end up with a chain with $N_a+3$ matter sites, with $N_{q}$ charged (or odd-length) clusters and $N_{0}$ neutral (or even-length) clusters. The odd (even) length clusters have a minimum extent of 1 (2), but since clusters have a minimum separation of 2 vacuum sites, for the purposes of counting we first add a vacuum site to the left and right-most edges of the chain, then combine every matter cluster with their immediate neighboring sites (necessarily vacuum), treating them as ``blocks'' of minimum extents 3 (for charged clusters) or 4 (for neutral clusters). This $N_a+5$ site chain thus consists of these minimal blocks in some order, inflated with additional pairs of matter sites within each block or pairs of vacuum sites between each block. Hence, we have the relation: $N_a+5 = 3N_{q}+4N_{0}+2k$, where $k$ is the number of inflated pairs. 

The dynamics of $H_{\text{LGT}}$ allows charge clusters to shrink and grow by 2 sites, or equivalently, allows the positions of the inflated pairs of sites to shuffle around the chain. Hence, the dimension of a Krylov fragment is exactly the number of ways that one can shuffle k pairs of inflated sites around the $2(N_{q}+N_{0})-1$ slots between or within blocks. We end up with:
\begin{equation}
    \text{dim}(N_a,N_{q},N_{0}) = \begin{pmatrix}
\frac{N_a+N_{q}+1}{2} \\
\\
\frac{N_a+5-3N_{q}-4N_{0}}{2}
\end{pmatrix}
\end{equation}
where the fragment dimension $\text{dim}(N_a,N_{q},N_{0})$ is a binomial coefficient. We shuffle the ordering of the $N_{q}$ and $N_{0}$ clusters and end up with a distinct fragment of identical dimension. 

Finally, summing over the possible numbers of charge clusters $\{N_{q},N_{0}\}$ gives us exact expressions for both the total number of Krylov fragments and the total dimension of the Hilbert space. They are respectively:
\begin{equation} \label{krylovnum}
    N_{\text{Krylov}} = \sum_{\{N_{q},N_{0}\}}\begin{pmatrix}
N_{q} + N_{0} \\
N_{q}
\end{pmatrix}
\end{equation}
\begin{equation} \label{hilbertdim}
    D_{\text{total}} = \sum_{\{N_{q},N_{0}\}}\text{dim}(N_a,N_{q},N_{0})\begin{pmatrix}
N_{q} + N_{0} \\
N_{q}
\end{pmatrix}
\end{equation}

As a sanity check, we verify that the expression \label{hilbertdim} does indeed evaluate to the Fibonacci number $F_{N_a+2}$, a well-known result for the Hilbert space dimension of the nearest-neighbor blockaded PXP model \cite{Lesanovsky_2012,moudgalya2022quantum}. A numerical plot of $N_{\text{Krylov}}$ (Extended Data Fig.~\ref{fig:figure6}a) readily shows that the number of Krylov fragments grows exponentially, as $N_{\text{Krylov}} \propto 1.22^{N_a}$. Moreover, the dimension of the largest fragment vanishes relative to the entire Hilbert space dimension (Extended Data Fig.~\ref{fig:figure6}b), verifying strong fragmentation in $H_{\text{LGT}}$. We additionally observe that the fraction of fragments that remain completely frozen appears to rapidly approach $1/3$ (Extended Data Fig.~\ref{fig:figure6}c).

\section*{Stability of fragmentation to perturbations}
We now briefly discuss the effect of high-order terms on the fragmentation in $H_{\text{LGT}}$. Referencing the second-order effective Hamiltonian (Eq.~(\ref{eq:secondorder})), the first term is diagonal in the $Z$-basis and leaves $Z$-product states invariant. It only results in energy offsets between states that are much smaller in magnitude than $\Omega$. Therefore, we do not expect this term to impact the fragmentation in $H_{\text{LGT}}$. 

For the second term, we note the following equalities:
\begin{widetext}
\begin{align}
    \left(P_{i-1}P_{i}\tilde{\sigma}^+_{i+1}P_{i+2}Q_{i+3}\right)\left(Q_{i-3}P_{i-2}\tilde{\sigma}^-_{i-1}P_{i}P_{i+1}\right) + \text{H.c.} =\; & Q_{i-3}P_{i-2}\tilde{\sigma}^-_{i-1}P_{i}\tilde{\sigma}^+_{i+1}P_{i+2}Q_{i+3} + \text{H.c.} \\
    \left(P_{i-3}P_{i-2}\tilde{\sigma}^-_{i-1}P_{i}Q_{i+1}\right)\left(Q_{i-1}P_{i}\tilde{\sigma}^+_{i+1}P_{i+2}P_{i+3}\right) + \text{H.c.} =\; & P_{i-3}P_{i-2}\tilde{\sigma}^-_{i-1}P_{i}\tilde{\sigma}^+_{i+1}P_{i+2}P_{i+3} + \text{H.c.}    
\end{align}
In other words, the second spin-exchange term in Eq.~(\ref{eq:secondorder}), which arises from virtual excitations outside the $V_1$ block, is equivalent to the sequential application of two bond-flips permitted by $H^{(1)}_{\text{eff}}$. This process preserves the block connectivity. Hence, we do not expect $H^{(2)}_{\text{eff}}$ to affect the fragmentation of $H_{\text{LGT}}$. 

Alternatively, in the LGT picture, the second term may also be written as:
\begin{align}
      \sum_{j=2}^{N-3} \Pi'_{j-3,j+2}\sigma^-_{j-2}S^+_{(j-2,j-1)}\sigma^+_{j-1}\sigma^+_{j}S^-_{(j,j+1)}\sigma^-_{j+1} + \text{H.c.}
\end{align}
\end{widetext}
This corresponds to a spin-exchange between next-nearest neighbor links and a dipole-moment conserving term on the matter sites. The projector is defined as: $\Pi'_{j,j+1} = \ket{\circ_j\bullet_{j+1}}\bra{\circ_j\bullet_{j+1}} + \ket{\bullet_j\circ_{j+1}}\bra{\bullet_j\circ_{j+1}}$. Under such a term, the positions of clusters may shift but the ordering of charged or neutral clusters remains invariant and the fragmentation is unaffected. Longer-range interaction terms are at most on the order of $V_2 \approx 2\pi \times 140$~kHz. Similar to the first term in $H^{(2)}_{\text{eff}}$, they result in additional small energy offsets of states much smaller than $\Omega$ and hence are not expected to reduce Hilbert space connectivity. 

Finally, we note that spin-exchange terms mediated by virtual excitations that violate nearest-neighbor blockade, such as terms of the form $P_{i-2}P_{i-1}\sigma^{+}_i\sigma^{-}_{i+1}P_{i+2}Q_{i+3}$, can break the fragmentation structure. However, these terms appear at second order in $\frac{\Omega}{V_0}$, with an amplitude of $\frac{\Omega^2}{4V_0}\approx 2\pi\times0.83$~kHz, several orders of magnitude smaller than $V_1$ and negligible for our experimental timescales.

\section*{Analytical expressions for infinite-temperature SLIOM distributions}
In computing infinite-temperature expectation values of SLIOM distributions (Fig.~\ref{fig:4}b,c), one may manually enumerate the cluster centers-of-mass for every $Z$-basis microstate and histogram them by sites. This approach becomes quickly intractable for larger system sizes, so it is useful to derive analytical expressions instead. We first derive an expression for the case of the left boundary SLIOM $q_1$. 

\begin{figure*}[!htbp]
    \centering
    \includegraphics[keepaspectratio,width=18cm]{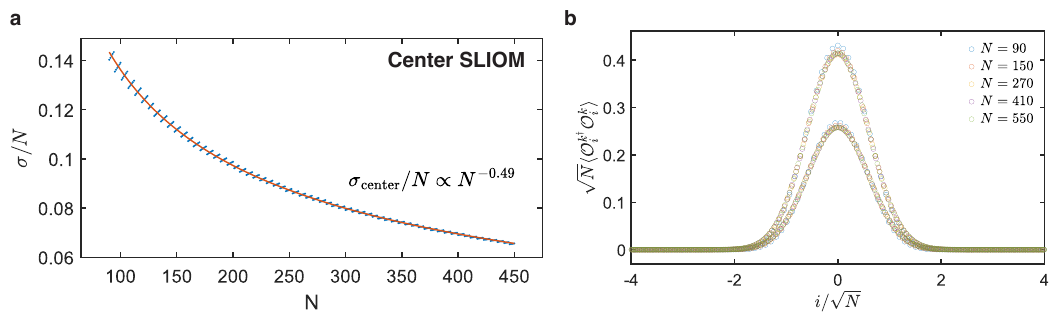}
        \caption{\textbf{Scaling properties of the center SLIOM in the largest symmetry sector.} \textbf{a}, Full-width-half-maxima $\sigma$ of the center SLIOM distributions over the system size $N$, evaluated for system sizes ranging from $N=90$ to $N=450$. The number of clusters in the largest symmetry sector does not necessarily increase with every increment in the system size, explaining the sawtooth feature in the plot that becomes less prominent for larger N. The fractional width is observed to scale with an exponent $\alpha = 0.49$, in agreement with our results in the main text. \textbf{b}, A scaling collapse of center SLIOM distributions shows that their width indeed appears to scale as $\sqrt{N}$. The site index $i$ is plotted such that the center site on the chain of particles corresponds to $i=0$. Each curve has a jagged feature between half-integer sites and integer sites, leading to separated Bell-shaped distributions.
}
    \label{fig:figure7}
\end{figure*}

Specifically, we find the probability $q_1(x)$ that the center-of-mass of the leftmost cluster is at site $x$, on a chain with $N_a+3$ matter sites. In an infinite temperature ensemble, this is simply given by the relative number of such configurations. We start by noting that if the leftmost cluster has an even (or odd) length, it occupies the first $2x$ (or $2x-1$) sites of the chain. The rest of the chain has $N_{a}+3-2x$ (or $N_a+3-2x+1$) sites. As before, we may add a vacuum site to both ends of the chain, shuffle inflated pairs anywhere in between and within minimal blocks, shuffle the ordering of odd and even length clusters, and vary the numbers of each type of cluster. We end up with the expression:
\begin{widetext}
\begin{equation} \label{boundary_analytical}
    q_1(x) = \frac{1}{D_{\text{total}}}\sum_{\{N_{q},N_{0}\}}\begin{pmatrix}
\frac{N_a+N_{q}-1}{2} + (2-x) \\
\\
\frac{N_a-3N_{q}-4N_{0}}{2} + (2-x)
\end{pmatrix}
\left[\begin{pmatrix}
N_{q} + N_{0} \\
N_{q} - 1
\end{pmatrix} + \begin{pmatrix}
N_{q} + N_{0} \\
N_{0} - 1
\end{pmatrix}
\right]
\end{equation}
Numerically evaluating $q_1(x)$ shows exact agreement with the distribution obtained from manual enumeration of microstates, for system sizes up to $N_a=30$, verifying its accuracy. It is similarly possible to derive expressions for the infinite-temperature distributions of bulk SLIOMs, although they are considerably more complex. 

Now, we compute $q_{k}(x)$, the relative number of configurations on a chain with $N_a+3$ matter sites, where the $k^{\text{th}}$ cluster from the left has a center-of-mass at site $x$. Again, we add a vacuum site on each end of the chain, and combine clusters with their immediate neighboring sites to form blocks. The length of the $k^{\text{th}}$ block may vary, and we denote it by $2s+1$. We therefore split the chain into three parts: the sub-chain to the left of the $k^{\text{th}}$ cluster (of length $N_l = x-\lceil s + 1/2\rceil$), the $k^{\text{th}}$ cluster, and the sub-chain to the right of the $k^{\text{th}}$ cluster (of length $N_r = N_a+5-x-\lfloor s + 1/2\rfloor$). We may then separately enumerate the total number of configurations for the two sub-chains with an expression similar to $D_{\text{total}}$, varying the length $2s+1$ of the $k^{\text{th}}$ cluster. The number of clusters on the left sub-chain is constrained to $N_c = k-1$ and we obtain the expression (for $k>1$):
\begin{equation} \label{bulk_analytical}
     q_k(x) = \frac{1}{D_{\text{total}}}\sum_{\{N_{q}^l,N_{0}^l\}}\sum_{\{N_{q}^r,N_{0}^r\}} \sum_{s=1}^{M}
     \begin{pmatrix}
\frac{N_l(s)+N_{q}^l+3}{2} \\
\\
\frac{N_l(s)-3N_{q}^l-4N_{0}^l}{2}
\end{pmatrix}
\begin{pmatrix}
N_{q}^l + N_{0}^l \\
N_{q}^l
\end{pmatrix} 
\begin{pmatrix}
\frac{N_r(s)+N_{q}^r+3}{2} \\
\\
\frac{N_r(s)-3N_{q}^r-4N_{0}^r}{2}
\end{pmatrix}
\begin{pmatrix}
N_{q}^r + N_{0}^r \\
N_{q}^r
\end{pmatrix} 
+    q_{k}^{r}(x)
\end{equation}
\begin{equation*}
    q_{k}^{r}(x) = \frac{1}{D_{\text{total}}}\sum_{\{N_{q}^l,N_{0}^l\}}\sum_{\{N_{q}^r,N_{0}^r\}}\left[\begin{pmatrix}
\frac{2x-N_a+N_{q}^l-3}{2} \\
\\
\frac{2x-N_a-6-3N_{q}^l-4N_{0}^l}{2}
\end{pmatrix} + \begin{pmatrix}
\frac{2x-N_a+N_{q}^l-2}{2} \\
\\
\frac{2x-N_a-5-3N_{q}^l-4N_{0}^l}{2}
\end{pmatrix}
\right]
\begin{pmatrix}
N_{q}^l + N_{0}^l \\
N_{q}^l
\end{pmatrix} 
\end{equation*}
\end{widetext}
where $N_{q}^l$ and $N_{0}^l$ ($N_{q}^r$ and $N_{0}^r$) count the number of blocks on the left (right) sub-chain with odd and even length blocks respectively. The sum over $N_{q}^l$ and $N_{0}^l$ is additionally constrained by the condition that there are $k-1$ blocks on the left sub-chain as $N_{q}^l + N_{0}^l = k-1$ and $s$ is summed to its maximal value of $M = \min (N-x,x-1)$. $q_{k}^{r}(x)$ is the additional contribution for the case that the $k^{\text{th}}$ block is the right-most block (where $N_r$ would be zero). While the constrained sums above make it challenging to obtain a closed-form expression, we again observe exact agreement with the distributions obtained from manual enumeration of microstates for all system sizes up to $N_a=30$. 

\section*{Scaling collapse of cluster distributions}
Enumerating the weighted average of all bulk SLIOM distributions as in Fig.~\ref{fig:4}d gets computationally challenging for larger system sizes, as one has to evaluate the distributions of an extensive number of charge clusters. To access even larger system sizes and to further support our scaling results obtained in Fig.~\ref{fig:4}d, we instead examine the distribution of just the center SLIOM in the largest symmetry sector. For example, the largest symmetry sector of an $N_a=200$ atom chain consists of $35$ charge clusters and the center SLIOM distribution corresponds to the distribution of the $18^{\text{th}}$ cluster from the left of the chain. We expect such distributions to peak at the center of the chain; as a reminder, we are interested in how the width (full-width-half-maximum) of this center distribution scales with the size of the system. 

We numerically evaluate these widths for system sizes ranging from $N_a=90$ to $N_a=450$ and plot them in Extended Data Fig.~\ref{fig:figure7}a. In agreement with the results in the main text, the fractional width scales approximately as $\sqrt{N}$. The sawtooth feature in the plot, where the fractional width rises and then drops discretely once roughly every six increments in $N_a$, is because the number of clusters in the largest symmetry sector increases by one only about every six increments in $N_a$. This may be crudely reasoned as follows: the minimum extent of a matter block is three sites, so the maximum number of blocks on a chain with $N_a+3$ matter sites is $\lfloor\frac{N+5}{3}\rfloor$, which represents a configuration saturated with blocks, without inflated pairs of sites. For a general chain, one can estimate that the number of ways to permute minimal blocks and inflated pairs (equal to the size of $\mathcal{H}_{N_c}$) is roughly maximal when the number of blocks is half of this maximal value, i.e.\  $\lfloor\frac{N+3}{6}\rfloor$. In practice, where we numerically identify the largest symmetry sector, this crude estimate remains surprisingly accurate. 

As a clearer visualization of the scaling, Extended Data Fig.~\ref{fig:figure7}b shows that the center SLIOM distributions for various system sizes are almost identical when their widths (heights) are rescaled by a factor of $1/\sqrt{N}$ $(\sqrt{N})$. Here, we see two Bell-shaped distributions for each system size, that are in fact part of the same overall jagged distribution. This feature occurs for the same reason as in Fig.~\ref{fig:4}c: charge clusters that are centered on half-sites are neutral clusters, which are on average one site longer than charged clusters. Hence, the projectors $O^k_i$ onto such configurations project onto a Hilbert space of lower dimension, resulting in a smaller peak. In fact, we see that the ratio of peak heights between the two Bell distributions for $N=550$ is about $1.61$, which is remarkably close to the ratio of total Hilbert space dimension for chains of length $N+1$ and $N$ in the limit of large $N$, i.e.\ $D_\text{total}(N+1)/D_\text{total}(N) \approx 1.618$.

\section*{Data availability}
The data that support the findings of this study are available from the corresponding author upon reasonable request.

\begin{acknowledgments}
We acknowledge earlier contributions to the experiment construction from Weikun Tian, Fan Jia, Wen Jun Wee, An Qu, and Jiacheng You, as well as helpful discussions with Mohammad Mujahid Aliyu. W.W.H.\ is supported by the 
National Research Foundation (NRF), Singapore, through the NRF Fellowship NRF-NRFF15-2023-0008, and through the National Quantum Office, hosted in A*STAR, under its Centre for Quantum Technologies Funding Initiative (S24Q2d0009). N.K.\ acknowledges funding in part from the NSF STAQ Program (PHY-1818914). 
\end{acknowledgments}

\section*{Author contributions}
P.R.D.\ and L.Z.\ ran the experiments and developed the theory simulations. W.W.H.\ and N.K.\ guided the theory work. H.L.\ supervised the project. All authors discussed the results and contributed to the manuscript. \\

\section*{Competing interests}
The authors declare no competing interests.

\section*{Materials and correspondence}
Correspondence and request for materials should be addressed to H.~Loh.

\end{document}